# Zinc oxide induces the stringent response and major reorientations in the central metabolism of *Bacillus subtilis*

Sylvie Luche[1], Elise Eymard-Vernain[1], Hélène Diemer[2,3], Alain Van Dorsselaer[2,3], Thierry Rabilloud[4], and Cécile Lelong[1*]

1 Pro-MD team, Université Joseph Fourier, CEA Grenoble, iRTSV, Laboratoire de Chimie et Biologie des Métaux, UMR CNRS-CEA-UJF, Grenoble

2 Laboratoire de Spectrométrie de Masse Bio-Organique, IPHC, Université de Strasbourg , 25 rue Becquerel 67087 Strasbourg, France

3 CNRS, UMR7178, 67087 Strasbourg, France

4 Pro-MD team, UMR CNRS 5249, Laboratoire de Chimie et Biologie des Métaux, UMR CNRS-CEA-UJF, Grenoble

*To whom correspondence should be addressed: cecile.lelong@univ-grenoble-alpes.fr

Running title: *Bacillus subtilis* responses to zinc oxide

**Summary**

Microorganisms, such as bacteria, are one of the first targets of nanoparticles in the environment. In this study, we tested the effect of two nanoparticles, ZnO and $TiO_2$, with the salt $ZnSO_4$ as the control, on the Gram-positive bacterium *Bacillus subtilis* by 2D gel electrophoresis-based proteomics. Despite a significant effect on viability ($LD_{50}$), $TiO_2$ NPs had no detectable effect on the proteomic pattern, while ZnO NPs and $ZnSO_4$ significantly modified *Bacillus subtilis* metabolism. These results allowed us to conclude that the effects of ZnO observed in this work were mainly attributable to Zn dissolution in the culture media. Proteomic analysis highlighted twelve modulated proteins related to central metabolism: MetE and MccB (cysteine metabolism), OdhA, AspB, IolD, AnsB, PdhB and YtsJ (Krebs cycle) and XylA, YqjI, Drm and Tal (pentose phosphate pathway). Biochemical assays, such as free sulfhydryl, CoA-SH and malate dehydrogenase assays corroborated the observed central metabolism reorientation and showed that Zn stress induced oxidative stress, probably as a consequence of thiol chelation stress by Zn ions. The other patterns affected by ZnO and $ZnSO_4$ were the stringent response and the general stress response. Nine proteins involved in or controlled by the stringent response showed a modified expression profile in the presence of ZnO NPs or $ZnSO_4$: YwaC, SigH, YtxH, YtzB, TufA, RplJ, RpsB, PdhB and Mbl. An increase in the ppGpp concentration confirmed the involvement of the stringent response during a Zn stress. All these metabolic reorientations in response to Zn stress were probably the result of complex regulatory mechanisms including at least the stringent response *via* YwaC.

## Introduction

In the past decade, nanoparticles (NPs) have been used more and more in research and in industry, which has raised questions about the potential toxicity of such materials. Nanomaterials are used in various fields including drug delivery, electronics, therapeutics, cosmetics, water treatment technology, environmental remediation and paint components. Due to these applications, many varieties of NPs are being industrially produced in massive amounts. These NPs are discharged into the environment during their life cycle or *via* nanomaterial-containing wastes. This waste or contamination may provide a path of entry into the food chain *via* microorganisms and/or eventually disturb the ecological balance. NPs are naturally present in the environment, e.g. in volcanic dusts in the atmosphere or as nanometer-scale particles from soil erosion. Ecosystems have evolved with these NPs for million years and organisms are adapted to them. By contrast, manufactured NPs released into the environment are new components that have arrived in significant amounts. Microorganisms, including bacteria present in the natural ecosystem, are among the primary targets that are exposed to NPs.

NPs are highly reactive and catalytic in nature compared to their ions or bulk counterparts (see [1] for review). They exhibit toxicity to living organisms mainly because of their small size, large surface-to-volume ratio and highly reactive facets. Because of their distinct properties compared to their corresponding ions or bulk material, NPs can lead to unpredictable effects on biological systems. $TiO_2$ and ZnO NPs are commonly used in sunscreens and other personal care products, paints, electronic devices and solar cells because of their optical, piezoelectric, magnetic and gas sensing properties, which leads to the release of these NPs into environment *via* industrial and domestic waste. Thus, it is very important to evaluate their environmental impact using one of the first pathways of entry into the ecosystem, bacteria. *Bacillus subtilis,* the Gram-positive bacterium, can be isolated from a wide range of environments, both terrestrial and aquatic, suggesting that this species is ubiquitous and broadly adapted to grow in diverse settings within the biosphere [2–4]. This bacterium is known to have a highly versatile metabolism, which can quickly and easily adapt to numerous environmental modifications. Every modification due to the presence of toxic elements such as NPs will impact its physiology [5–7]. This property is one of the reasons explaining why this bacterium has been found in many biotopes in which it has genetically evolved to obtain the best possible fitness. One of the natural biotopes of this bacterium is the upper layer of soil where it can modify the growth of some vegetal species by modulating the carbon and nitrogen cycles surrounding the roots. It has been isolated from the rhizosphere of a variety of plants in greater numbers than most other spore-forming bacteria [8–11]. Considering that *Bacillus subtilis* is found on and around plants and that many animals consume plants, it is no wonder that this bacterium is often found in gut or in feces [12–15]. The passage of *Bacillus subtilis* through the gastrointestinal tract may not be without effect. However, *Bacillus subtilis* is a ubiquitous organism, not solely a consequence of spore persistence in these environments, but also because of its capacity to grow in diverse environments including soils, plant roots and the gastrointestinal tract of animals. The presence of NPs in the environment in an increased concentration will have significant consequences on *Bacillus subtilis* metabolism, which will also impact the physiology of surrounding organisms in the plant and animal kingdoms.

Most of the analysis performed on the effect of NPs on bacteria have only studied the antibacterial properties of NPs [16–19]. Some studies have also shown effects on membrane disorganization [20], the production of reactive oxygen species (ROS) [21–23] or the production of exopolysaccharide [24], as well as altered enzyme expression profiles [25]. We are the first to study the metabolic effects of NPs on *Bacillus subtilis* using a proteomic approach. Despite a significant effect on viability, $TiO_2$ NPs had no detectable effect on the proteomic pattern, while ZnO NPs and $ZnSO_4$ salt significantly modified central metabolism and the stringent response of *Bacillus subtilis*.

## Experimental procedures

### Bacterial strain and culture media

The *Bacillus subtilis* strain used was the 168 strain (*trpC2*, wild type) (personal gift, Maria Laaberki). The medium was LB: 10 g/l tryptone, 5 g/l yeast extract and 5 g/l NaCl supplemented with 12 g/l agar when necessary. Cells were grown in Erlenmeyer flasks with shaking at 200 rpm at 37°C.

The nanoparticles were from SIGMA: $TiO_2$ (ref 700347, mixture 80:20 of anatase and rutile, 33-37 wt.% in water, <150 nm volume distribution, DLS, (dynamic light scattering)), and ZnO (ref 721077, 50% in water, <100 nm by DLS) were already purchased as water dispersions. Their principal features have been described elsewhere [26] and [27] in press.

### Nanoparticles stresses

Cells were grown in LB medium (250 ml) in Erlenmeyer flasks. When the $OD_{600}$ reached 0.6, 50 ml of the culture was transferred to a new Erlenmeyer flask, and the cells were incubated without for the control and with 130 μg/ml $TiO_2$, or 170 μg/ml ZnO or 0.3 mM $ZnSO_4$ for one hour under shaking at 37°C. Then, aliquots (100 μl) of the samples were diluted in $H_2O$ and appropriate dilutions were plated onto LB agar, and incubated overnight at 37°C. The CFUs were counted and the viability was determined. At the same time, the remaining 50 ml was harvested by centrifugation at 5000 rpm for 15min at 4°C. The pellet was washed twice with 50 ml of PBS and then frozen at -20°C until used. For the metabolic assays, the sample was separated in 2 or 10 ml aliquots before harvest by centrifugation. The pellet was frozen in liquid $N_2$. All experiments were performed in triplicate and at least two technical replicates from each dilution were plated to determine the number of CFUs.

### DLS measurements

Nanoparticles were diluted in three different solutions just before measurement with the DLS (dynamic light scattering) apparatus (Wyatt technology): $H_2O$, LB medium or preconditioned LB medium. The preconditioned LB was prepared as follows: *Bacillus subtilis* strain 168 was grown in LB until an $OD_{600}$ of 0.6. Then, 2 ml of culture was harvested by centrifugation at 13,000 rpm for 10 min. The supernatant was filtered through a 0.2 μm filter. The filtered solution constituted preconditioned LB. Each measurement was performed at least in triplicate.

### Sample preparation of the cytosolic crude extract for 2D-gel electrophoresis experiments

The frozen pellets were resuspended and incubated in 225 μl of lysis solution containing 200 μl of TES buffer (10 mM Tris, 1 mM EDTA, 0.2 M sucrose) + 20 μl of 30 mg/ml lysozyme, and 5 μl of 1/50 benzonase solution for 20 min at 37°C. The samples were sonicated in a water bath for 5 min. Next, 6 μl of 20% SDS and 80 μl of 1 M DTT were added before incubation at 100°C for 5min. The lysis was achieved by sonication in a water bath for 5 min. Finally, 800 µl of rehydration solution (7 M urea, 2 M thiourea, 4% CHAPS, 0.4% carrier ampholytes (Pharmalytes 3-10) and 100 mM dithiodiethanol [28,29] was added. Each sample was quantified after separation by SDS-PAGE and staining: known quantities (10, 20 and 30µg) of total protein extracts and two dilutions of cytosolic crude extract extracts were separated on the same gel. Proteins were stained with colloidal Coomassie blue [30] (data not shown).

**2D-gel electrophoresis**

IEF: Homemade 160 mm long 4-8 or 3-10.5 linear pH gradient [31] gels were cast according to published procedures [32]. Four millimeter-wide strips were cut and rehydrated overnight with the sample, diluted in a final volume of 0.6 ml of rehydration solution and 100 mM dithiodiethanol [28,29].

The strips were then placed on a Multiphor plate and IEF was carried out with the following electrical parameters: 100 V for 1 hour, then 300 V for 3 hours, then 1000 V for 1 hour, then 3400 V up to 60-70 kVh. After IEF, the gels were equilibrated for 20 minutes in 125 mM Tris, 100 mM HCl, 2.5% SDS, 30%glycerol and 6 M urea. Gels were then transferred on top of SDS gels and sealed with 1% agarose dissolved in Tris 125mM, HCl 100mM, SDS 0.4% and 0.005% (w/v) bromophenol blue.

SDS electrophoresis and protein detection: 10% T gels (160x200x1.5 mm) were used for protein separation. The Tris taurine buffer system was used at an ionic strength of 0.1 and a pH of 7.9 [33]. The final gel composition was 180 mM Tris, 100 mM HCl, 10% (w/v) acrylamide and 0.27% bisacrylamide. The upper electrode buffer was 50 mM Tris, 200 mM taurine and 0.1% SDS. The lower electrode buffer was 50 mM Tris, 200 mM glycine and 0.1% SDS. The gels were run at 25 V for 1 hour, then at 12.5 W per gel until the dye front had reached the bottom of the gel. Detection was carried out by fast silver staining [34].

Image analysis: Image analysis was performed with Delta2D software (v.3.6) (Decodon, Germany). Briefly, three gel images arising from three different cultures and protein preparations were warped for each group onto a master image, one for the culture in LB (control without nanoparticles or salt) and one for each stress condition ($TiO_2$, ZnO nanoparticles and $ZnSO_4$ salt). The control master gel image was then warped onto each stress condition master gel image and a union fusion image of all the gel images was then made. Detection was carried out on this fusion image, and the detection results were then propagated to each individual image.

The resulting quantification table was then analyzed using the Student's t-test function in the software, and the spots having both a p-value lesser than 0.05 and an induction/repression ratio of 1.5 or greater were selected for further analysis by mass spectrometry after being manually verified. For the global analysis of the power and reproducibility of the experiments, the Storey and Tibshirani approach was used [35], as described by Karp and Lilley [36] (Supplementary Figure S2).

The spots of interest were excised from a silver staining gel by a scalpel blade and transferred to a 96-well microtitration plate. Destaining of the spots was carried out by the ferricyanide-thiosulfate method [37] on the same day as silver staining to improve sequence coverage in the mass spectrometry analysis [38].

**Mass spectrometry**

Analysis was carried out essentially as described previously [39], except that only the nanoLC-QTOF-MS system was used for nanoLC-MS/MS analysis and the version of the UniProtKB/SwissProt database was 2012_12 (538,585 sequences) for protein identification. The detailed protocol can be found in the supplementary data (S1).

**Crude extract preparation for enzymatic and biochemical (SH, CoA, ppGpp) assays**

Cells pellets were obtained from 10 ml aliquots of *Bacillus subtilis* strain 168 cultures grown with or without stress, as described above. Cells were incubated in a 200 µl final volume of 20 mM phosphate buffer (pH 7.4) containing 15 µl of 30 mg/ml lysozyme and 4 µl of 1/50 benzonase solution for 20 min at 37°C. This was then centrifuged at 13,000 rpm at 4°C for 10 min. Total protein quantification was performed using the Bradford assay.

**Malic enzymes activity assay**

NAD(P)H malic enzyme assays were performed as previously described [40]. Briefly, crude extract supernatants were obtained as described above and stored on ice until malic enzyme assays were performed. Total protein quantification was performed using the Bradford assay.

Malic enzyme activities were measured by spectrophotometrically monitoring at 340 nm, i.e. NAD(P)H formation during the reductive decarboxylation of malate (malate+ NAD(P)=pyruvate+ NAD(P)H). The reaction mixture was composed of 10 µl of an appropriate dilution of crude extract supernatant in 1 ml of buffer (50 mM Tris HCl pH 8, 10 mM $MgCl_2$, 10 mM $MnCl_2$, 50 mM KCl and 10 mM thioglycerol). The reaction was started by the addition of 5 mM of $NAD^+$ or $NADP^+$. The specific activity was calculated using the initial rate data normalized by the total protein quantity.

**SH assay**

Crude extracts were obtained as described above. Aliquots of the crude extract were used to measure the total protein quantity using the Bradford assay and the total sulfhydryl groups (total SH). To measure the free sulfhydryl groups, proteins of the crude extracts were precipitated with 10% TCA for one hour on ice, harvested by centrifugation (13,000 g, 5 min). Free SH was measured in the supernatant. SH concentrations were determined spectrophotometrically as follow: 20 µl of the crude extract or 100 µl of the supernatant were mixed in 1 ml of 30 mM Tris HCl pH 8, 3 mM EDTA, 1.5 mM DTNB and 40% ethanol, then incubated one hour at room temperature before measuring the optical absorbance at 412 nm.

**CoA assay**

Coenzyme A concentrations were determined using the Coenzyme A assay kit from Sigma-Aldrich; 40 µl of crude extracts were used for each tested condition. CoA concentrations were normalized by the protein concentrations measured by the Bradford assay.

**Catalase activity assay**

A 10 ml volume of bacteria grown until 0.6 $OD_{600}$ were stressed with 130 µg/ml $TiO_2$, 170µg/ml ZnO, 0.3mM $ZnSO_4$ or 5mM $H_2O_2$ for one hour under shaking at 37°C. Bacteria were then pelleted by centrifugation at 6,000 rpm for 10 min. The pellets were washed with 1 ml of PBS and centrifuged (13,000 rpm, 5 min). The pellets were finally resuspended in 8 ml of PBS. The catalase activities were immediately determined on cells using a Clarke-type electrode (Hansatech, Kings Lynn, UK), as described by Rorth and Jensen [41]. The specific activities were calculated as the number of generated $O_2$ moles/min/CFU.

**Inductively coupled plasma atomic emission spectroscopy (ICP-AES)**

Shimadzu ICP 9000 (with Mini plasma Torch in axial reading mode) was used to measure the zinc content. Standard solution for AAS (Sigma Aldrich) was used to generate the calibration curve between 10 to 500 µg/L in pure water with 1% of HNO3 (Fluka). Samples were routinely incubated in HNO3 10% ON at RT. Briefly, 500µl of each cytosolic crude extract was incubated with 90µl of HNO3 65% ON at RT. The sample was centrifuged at 13,000 rpm for 5 min. Before to be measured pure water was added to the supernatant extemporaneously to obtain a final volume of 7 ml. Ytterbium solution standard for AAS (Sigma Aldrich) was used as an internal standard to prevent calibration drift and fluidic perturbation. The result was express in µg/L [Zn] per mg of protein.

**ppGpp assay**

The ppGpp content was measured as previously described [42], using an on-off fluorescent probe. ppGpp extraction was performed on crude extracts of *Bacillus subtilis* obtained as described above (growth in LB medium as the control and LB medium with a stress agent (130 µg/ml $TiO_2$, 170 µg/ml ZnO, 0.3 mM $ZnSO_4$ or 1% norvaline). Aliquots of the crude extract were used to measure the total protein quantity using the Bradford assay. The same quantity of each crude extract was precipitated with 10% formic acid for one hour one ice and cleared by centrifugation. The pH of the supernatant was adjusted to 7.4 by the addition of N-ethylmorpholine (130 µl for a sample of 200 µl).

Fluorescent probe preparation: 3 µM of oligonucleotide DNA (5’-GGC AGG TTG GGG TGA CTA AAA ACC CTT AAT CCC C-3’) and 90 µl of freshly prepared 1 mM silver lactate (Fluka BioChemika) were first mixed in 20 mM phosphate buffer (pH 6.6) for 5 min in a final volume of 1 ml, and kept protected from light at room temperature. After 20 min, 90 µl of freshly prepared 1 mM $NaBH_4$ (Sigma-Aldrich) was quickly injected, thoroughly mixed, and the mixture was kept in the dark without further disturbance at 4°C overnight.

Fluorescent detection: 20 µl of ppGpp extract was quickly mixed with the fluorescent probe (at a final concentration of 300 nM) in 20 mM phosphate buffer pH 7.4 in a final volume of 200 µl. The mix was incubated 45 min at room temperature and protected from light before the first measurement ($F_0$). Then, $Cu^{2+}$ was added at a final concentration of 800 nM and the mix was incubated 45 min at room temperature and protected from light before the second measurement (F). Fluorescent spectra were measured with a Tecan Infinite® M1000 instrument at a fluorescent excitation wavelength of 585 nm and fluorescent emission wavelength of 635 nm. The results show the average of a technical duplicate for each biological triplicate and are expressed as follow: $[F/F_0]$stress condition normalized by $[F/F_0]$without stress.

## Results

### NPs sizes in growth medium and cell viability

Nanoparticles suspensions of $TiO_2$ and ZnO were diluted in $H_2O$, LB medium or preconditioned LB medium before measurements with a DLS apparatus (Figure 1a,b). DLS measurements revealed that the different media modified the diameter of both nanoparticles. The $TiO_2$ NP diameter was increased in LB medium, preconditioned or not, compared to their size in $H_2O$. Nevertheless, the preconditioned medium decreased the diameter of $TiO_2$ NPs compared to the LB medium. ZnO NPs diameter was decreased in both LB and preconditioned LB compared to the $H_2O$ dilution, which may be explained by partial dissolution of the NPs [43]. DLS measurements of $TiO_2$ and ZnO suspensions in the different media performed over time (up to one hour) after mixing showed no significant differences with the initial values (data not shown).

Different dilutions of $TiO_2$ and ZnO NPs (from $10^{-1}$ to $10^{-4}$ mg/ml) and $ZnSO_4$ (from 0.03 to 0.5 mM) were investigated on *Bacillus subtilis* growth (Figure S3). 130 µg/ml $TiO_2$, 170 µg/ml ZnO and 0.3 mM $ZnSO_4$ were chosen because at these concentrations no major alteration of the growth curves. In these conditions, the viability levels were around 50% after one hour of treatment (Figure 1c). Such viability allowed for obtaining sufficient numbers of stressed, but not dead, cells to perform 2DE analysis.

### Differential patterns of protein synthesis in the presence of $TiO_2$ and ZnO NPs and $ZnSO_4$

In order to investigate the effects of NPs on the protein pattern, 2D protein gels of equal amounts of protein crude extract, from three independent cultures, were carried out. The gels were silver stained and analyzed with Delta 2D software: protein patterns from the crude extracts obtained after growth in LB medium were compared with the protein patterns of crude extracts obtained after growth in the presence of NPs or $ZnSO_4$. Despite a mortality of 40% after the addition of 130 µg/ml of $TiO_2$ for one hour, no significant difference was detected between the protein pattern from bacteria grown in LB medium and $TiO_2$ stressed bacteria (Figure 2a,b). In contrast, Delta 2D software analyses of the protein patterns of cells grown in the presence of ZnO NPs or $ZnSO_4$ highlighted 23 spots (Figure 2a,c,d), which were differentially expressed by a factor equal to or greater than 1.5 with a p-value lower than 0.05 in a two-tailed t-test. With the exception of YqjI, all highlighted spots showed the same profile in the presence of ZnO NPS and $ZnSO_4$ (Table 1). The proteins were identified by mass spectrometry. Proteins affected by ZnO NPs or $ZnSO_4$ stress were assigned to two main classes: proteins related to central metabolism and proteins related to the stringent response.

### Metabolic reorientations

Pentose pathway and TCA shift: twelve modulated proteins were related to central metabolism: MetE and MccB are related to cysteine metabolism, OdhA, AspB, IolD, AnsB, PdhB and YtsJ to the Krebs cycle and XylA, YqjI, Drm and Tal to the pentose phosphate pathway (PPP) (Figure 3). The two main enzymes involved in the oxidative branch and the non-oxidative branch of the PPP, YqjI and Tal, respectively, were increased after the addition of ZnO NPs or $ZnSO_4$. The non-oxidative branch converts ribose-5-phosphate back to glucose-6-phosphate, the substrate for the oxidative branch. The Tal enzyme, which catalyzes the reversible transfer of a dihydroxyacetone moiety derived from fructose-6-phosphate to D-erythrose-4-phosphate, forming D-sedoheptulose-7-phosphate and releasing D-

glyceraldehyde-3-phosphate, is a rate-limiting enzyme in the non-oxidative branch of the pathway in eukaryotic cells [44]. *Via* the oxidative branch, the pentose phosphate pathway is responsible for the generation of NADPH, which contributes to cellular defenses against oxidative stress. YqjI, which is the principal 6-P-gluconate dehydrogenase [45], was the only spot showing different profiles with ZnO NPs and $ZnSO_4$; its intensity increased in the presence of ZnO and was not significantly modified by $ZnSO_4$. The second main metabolic effect of ZnO NPs and $ZnSO_4$ was a reorientation of the Krebs cycle; the enzymes YtsJ, AspB and AsnB were significantly increased while OdhB was decreased. The activity of YtsJ, a malate dehydrogenase, described to function with both cofactors NAD or NADP [40,46], measured in the different crude extracts, was increased in all stressed conditions (Figure 4a,b). These modifications in expression lead to a partial shunt of the Krebs cycle in the part from $\alpha$-ketoglutarate to malate, in a way to increase the production of NAD(P)H and to release free CoA-SH in response to stress induced by ZnO NPs or $ZnSO_4$ (Figure 3 and Figure 4a,b).

Thiol metabolism: MetE and YrhB enzymes are involved in cysteine metabolism in opposite ways: MetE catabolizes the synthesis of methionine and YhrB catabolizes the synthesis of cysteine (Figure 5). Their expressions were also modified in opposite ways in the presence of ZnO NPs or $ZnSO_4$ salt: YrhB was increased and MetE was decreased (Figure 5, Table 1). The modifications in MetE and YrhB expression and the reorientation of central metabolism let us infer an increase in cysteine and CoA-SH concentrations in the cell under stress conditions (ZnO NPs and $ZnSO_4$ salt). The CoA-SH concentration on one hand and the ratio of free sulfhydryl groups/protein-bound sulfhydryl groups on the other were therefore measured in the presence or absence of NPs or Zn salt stress (Figure 4c,d). The proportion of free sulfhydryl was significantly increased in the presence of ZnO NPs and $ZnSO_4$, whereas the CoA-SH concentration was increased in all stress conditions, ZnO and $TiO_2$ NPs and $ZnSO_4$. *Bacillus subtilis* possesses three main low molecular weight (LMW) thiols, CoA-SH, cysteine and bacillithiol. Ma and colleagues [47] showed that bacillithiol is the major buffer of labile zinc in *Bacillus subtilis,* so most of the increase in free thiol in the presence of ZnO NPs and $ZnSO_4$ was probably the result of an increase in bacillithiol. The increasing of free intra cellular thiols following a stress with ZnO or Zn salt, was probably the consequence of an increasing of intracellular [Zn], in a concentration exceeding the physiological concentration. The intra cellular [Zn] was measured in all conditions. Figure 6b shows that in presence of Zn, nanoparticles or salt, the intracellular [Zn] was increased by four to six-fold, respectively.

The LMW thiols, in particular bacillithiol, are often described to play roles in the protection of cells against a variety of stresses, including oxidative stress [48–52]. To test the hypothesis of oxidative stress induction by ZnO NPs or $ZnSO_4$, catalase activity was measured in all conditions. Figure 6a shows that catalase activity was unchanged in the presence of $TiO_2$ NP, but increased by two-fold in the presence of ZnO, by four-fold in the presence $ZnSO_4$ and by three-fold in the presence of $H_20_2$ compared to the levels in control bacteria grown in LB medium. Taken together, the protective role of LMW thiols, the probable increase in NAD(P)H and the increased activity of catalase suggest that ZnO NPs and $ZnSO_4$ induce oxidative stress and/or thiol chelation stress in *Bacillus subtilis*.

**Stringent response**

The other main pattern affected by ZnO and $ZnSO_4$ was the stringent response. Nine proteins involved or controlled by the stringent response had their expression profile modified in the presence of ZnO NPs or $ZnSO_4$: YwaC, SigH, YtxH, YtzB, TufA, RplJ, RpsB and Mbl (Figure 2, Table 1). SigH is a sigma factor that controls the transcription of many genes involved in the transition from exponential growth to the stationary phase, including the initiation of spore formation and entry into the genetic competence state. Tam Le and collaborators [53], Eyman and coll [54] have shown that sigmaH activity and the stringent response are linked.

YwaC is a GTP pyrophosphokinase that catalyzes the synthesis of ppGpp, the alarmone produced during the stringent response in response to number of nutritional or environmental stresses [55–58]. *Bacillus subtilis* possesses three ppGpp synthases: RelA, YjbM and YwaC. RelA is a bifunctional enzyme that synthesizes and hydrolyses ppGpp, while YjbM and YwaC are only ppGpp synthases. In the presence of ZnO NPs or $ZnSO_4$, the expression of YwaC was increased by seven-fold. At the same time, the expression of several proteins known to be negatively or positively regulated by the stringent response was modified: YtxH, TufA, RpsB were increased and YtzB, RplJ and Mbl were decreased. YjbM, which presents a similar pI and MW as YwaC, was not detected among the spots as having undergone a modification in expression. As YwaC is a ppGpp synthetase, the ppGpp concentration was measured in the different crude extracts. In the presence of ZnO NPs, $ZnSO_4$ and norvaline, the proportion of ppGpp was significantly increased (by 1.8, 2.1 and 1.8-fold, respectively) while it was unchanged in the presence of $TiO_2$ (Figure 6b) compared to the cells grown in LB medium without any stress.

## Discussion

The proteomic analysis of *Bacillus subtilis* in the presence of nanoparticles or of the corresponding metallic ions showed that, despite 40% mortality, $TiO_2$ NPs led to no significant modification in the protein pattern, while ZnO NPs and $ZnSO_4$ led to important modifications in cell physiology. $TiO_2$ had no detectable effect on the *Bacillus subtilis* proteome, but although the free SH ratio was unchanged, Figures 4c and 4d show a significant increase in CoA-SH and sporulation after $TiO_2$ incubation, respectively. These observations led us to suppose that $TiO_2$ nanoparticles induce a stress which is not an oxidative stress leading to an increased expression of catalase, but to which cells probably respond *via* the activation of sporulation and of some parts of the oxidative stress pathway response, through not in a way detectable *via* 2D gels analysis.

Oppositely ZnO nanoparticles and $ZnSO_4$ salt induce marked responses in bacteria: the opposite response to that caused by $TiO_2$ considering obvious stress with metabolic reorientation and activation of the stringent response. The presence of Zn in the form of NPs or salt induced reorientation of central metabolism in a way to protect the cell by the production of NAD(P)H and free thiols. The diversion of methionine sulfur to cysteine is hypothesized to increase the concentration of low-molecular weight biothiols, in particular bacillithiol, which serves to maintain the reduced environment of the cytoplasm and/or to generate mixed disulfides in order to protect protein thiols from irreversible oxidation [59]. Moreover, cysteine is a substrate for the synthesis of CoA and very likely a substrate for the production of bacillithiol [50]. ICP assays have shown that [Zn] intracellular was significatively increased after a stress with the ZnO nanoparticles or Zn salt. The effects observed with Zn led us to suppose that ZnO nanoparticles are dissolved, the Zn salt enter into the cell and thus disturb cell homeostasis. Zn is redox-inert but has a high affinity for the side chains of Cys, Asp, Glu and His. Naturally, Zn binds proteins at structural sites or at the catalytic site of enzymes. When metal ions exceed their physiological concentration, they bind to enzymes, leading most of the time to their inhibition [60,61]. In most of the studied enzymes, Zn bound to cysteine thiols, which are involved in the structural or catalytic site. This binding to cysteine has two consequences on cell metabolism: first, it modifies the balance between free cysteines and disulfide bridges involved in combating oxidative stress [60,62] and second, it impedes central metabolism by the inactivation of several enzymes belonging to the TCA, glycolysis, the pentose phosphate pathway or respiration. In hepatocytes, aconitase, alpha-ketoglutarate dehydrogenase, isocitrate dehydrogenase-($NAD^+$ dependent), succinate dehydrogenase and cytochrome C oxidase are severely inhibited by Zn [63]. In *Saccharomyces cerevisiae*, Zn accumulation affects the correct [Fe–S] cluster assembly of aconitase, leading to inactivation of the enzyme [64]. In *Escherichia coli*, Zn ions strongly inhibit succinate dehydrogenase [65] and modulate the activity of several enzymes of central metabolism, e.g. enolase, cysteine synthase, serine hydroxymethyltransferase, phosphoglyceromutase, malate dehydrogenase and glyceraldehyde-3-phosphate dehydrogenase [66]. In some reports, the role of Zn ions has been correlated with modifications in detoxification activity by glutathione or modulation of thiol-related enzymes such as thioredoxin reductase. In *Bacillus subtilis*, Ma and collaborators [47] and Eide [67] showed that an excess of Zn is mainly buffered by low molecular weight bacillithiol. In our work, we observed an increasing production of free thiols and a reorientation of the Krebs cycle. In this latter pathway, the following enzymes: α-ketoglutarate dehydrogenase, succinyl-CoA synthetase and succinate dehydrogenase were shunted in the presence of Zn NPs or ions. Zn ions are already known to inhibit two of these, α-ketoglutarate dehydrogenase and succinate dehydrogenase, in eukaryotes and prokaryotes. To circumvent this problem, bacteria used the amino acids present in the LB medium to shunt the inhibited steps. AspB is used to convert aspartate and α-ketoglutarate into glutamate and oxaloacetate. AnsB is used to produce fumarate directly from aspartate. Moreover, as Zn also inhibits pyruvate production from glycolysis, YtsJ produces the necessary pyruvate from malate originating from the AsnB-FumC pathway. This shunt leads to an increasing concentration of NADH and free CoA-SH. Both NAD(P)H and free thiols (CoA-SH, Cys and bacillithiol) are known to be produced in response to oxidative stress. Even though no specific protein of oxidative stress, such as PerR or OhrR, was detected by the proteomic approach, metabolism reorientation (Krebs cycle and Cys production) seems to be the result of oxidative stress, which is the consequence of the modification of the free thiol homeostasis in response to the presence of an excess of ZnO and $ZnSO_4$. However, as CoASH and NADP(H) are also involved in fatty acid metabolism and as FabL was found to be regulated in the proteome, we can not rule out an effect on fatty acid metabolism.

The second main pattern affected by the presence of ZnO and $ZnSO_4$ was the induction of the stringent response *via* the ppGpp synthase YwaC. The production of the ppGpp "alarmone" may be the result of the activity of RelA, and/or YwaC and/or YjbM [55–58]. ppGpp is synthetized by the cell following nutrient starvation to switch off all reactions appropriate for growing cells such as ribosome synthesis, but it is also involved in the induction of RpoS, which controls the general stress and starvation response in Gram-negative bacteria [68]. In most Gram-positive bacteria, three genes coding for ppGpp synthase have been found. The most studied is the *relA* gene coding for the bifunctional protein RelA, a ppGpp synthase and hydrolase enzyme. The role and condition of expression of YjbM and YwaC is less known. YwaC has been shown to induce the transcription of *yvyD via* the sporulation-specific factor SigH. Proteomic analysis of the stringent response using a *relA* mutant and DL-norvaline to induce the stringent response [69] has shown that RplJ, TufA, RpsB and Mbl are negatively regulated by the stringent response while YtzB and YtxH are positively regulated by the stringent response. All of these regulations are at least partially under the control of RelA [69]. Under our conditions, the expression of YtzB, TufA and RpsB was modified in the opposite way compared to the data obtained by Eymann and colleagues [69]. YtxGHI is also under the control of $\sigma^H$. However, ppGpp is the main player of the stringent response. Gram-positive bacteria have at their disposal three ppGpp synthases, which are probably differentially expressed in response to various stimuli. YwaC has already been shown to be activated under salt, heat and EtOH stresses, independently of the stress response regulator SigB [70]. The induced expression of YwaC in a triple ΔrelAΔywaCΔyjbM mutant caused growth arrest in LB medium [57], probably due to a decrease in intracellular GTP, while the expression of YjbM in the same mutant background had no effect. Moreover, the expression of YwaC in the triple mutant drastically modified the transcriptome profile of the cell. All these observations led us to suppose that YwaC has a particular and fundamental role when *Bacillus subtilis* has to combat an environmental stress. The cell activates the stringent response to survive not only to amino starvation but also to chemical or physical stresses. Interestingly, it is more and more apparent that the stringent response is involved in the regulation of stress as well as in virulence or pathogenic processes [71–74].

Gram-positive bacteria can activate different pathways to respond to stress: the general stress pathway, the stringent response or the activation of sporulation (for sporulating bacteria). In the case of ZnO nanoparticles and $ZnSO_4$, the expression of SigH was decreased. Spo0H is the first sigma factor in a gene expression cascade involving a total of five factors, including spo0A, which leads to the formation of heat-resistant spores in *Bacillus subtilis* [54]. Spo0A is repressed by the stringent response *via* RelA [69]. Contradictorily, Ochi and colleagues [75] have shown that sporulation is induced by the stringent response, probably as an indirect effect due to the decrease in GTP. Mirouze and colleagues [76] have shown that the triple ppGpp mutant exhibits a delay in sporulation. Clearly, ppGpp plays a central role in the induction of sporulation, which may depend on the growth conditions. But more than its presence (or absence), its concentration and in parallel GTP availability fine-tune sporulation and the cell response to stress. CodY and stringent response are intimately linked in different Gram-positive bacteria [58,77–79]. In *Staphylococcus aureus*, Schoenfelder and colleagues [80] have shown that CodY and the stringent response regulate methionine biosynthesis. Considering the complex roles of CodY, the stringent response, and SigH in the response to stress demonstrated by proteomic analysis, we could speculate that the metabolic reorientations observed in response to Zn stress are the result of a complex regulation mechanism including at least the stringent response *via* YwaC, the SigH regulator and perhaps CodY.

Here, we have demonstrated that the presence of ZnO nanoparticles induces a stress that bacteria fight *via* large modifications of the metabolism. Similar observations have already been described by Hsueh and collaborators [81], who have demonstrated that ZnO nanoparticles affect the viability of *Bacillus subtilis* through the inhibition of cell growth, cytosolic protein expression and biofilm formation. Our study is a first step to understand, at a physiological level, the long-term adaptation of a gram-positive bacteria in response to nanoparticles. *Bacillus subtilis* is an ubiquitous bacteria which can grow in very different environments such as the upper layer of soil, marine sediment or mammalian gut. In all these places, *Bacillus subtilis* can adapt very rapidly its metabolism to survive to strong modifications of its environment [7] and it also interacts with, and contributes, to the growth of numerous biological organisms *via* the secretion of biopolymers e.g. lipoprotein, protease or DNA in the rhizosphere [82]. Nanoparticles are more and more used in industrial process and in everyday consumer products (solar screen, paint or medicine) [83,84]. This uncontrolled consumption induces an increasing amount of nanoparticles released in the environment. Nanoparticles contaminating the soil have direct cytotoxic, phytotoxic and genotoxic effects on plant physiology [85–88]. Modification of the bacteria metabolism also impacts indirectly the plant physiology via the plant-bacteria interaction. Moreover, the physico-chemical characteristics of the medium, as the soil for the plant, influence the effect of nanoparticles on its biological surrounding [89]. For example, nanoparticles have a non-negligible effect on the symbiotic process necessary for an optimal growth of plants by modifying, for example, the exchange of metal at the root level [90,91] or nitrogen fixation, or the root structure [92]. Understanding the effect of nanoparticles on bacteria physiology leaving in symbiosis with plants will allow thinking ahead potential consequences of such contamination on the evolution of the ecosystem [93] and will give forceful arguments for a better management of nanoparticles wastes.

## Acknowledgments

The financial support of the CEA toxicology program (Nanobiomet and Nanostress grants) is gratefully acknowledged. The support of the Labex SERENADE (11-LABX-0064) is also acknowledged. Finally, Dr. Caroline Barette and Dr. Julien Perard are acknowledged for their assistance in performing the fluorometric ppGpp assays and ICP-AES experiments, respectively.

## Table 1: Differentially-expressed proteins identified in the proteomic screen

| Protein | | | | | | Known regulon | Mass spectrometry Data | | Delta 2D Data | | | |
|---|---|---|---|---|---|---|---|---|---|---|---|---|
| N° | Symbol | UniProt accession number | function | PM (Da) | pI | | nb peptides | Percentage sequence coverage | ratio mean % Volume ZnO/ mean % Volume LB | t-test mean % Volume ZnO / mean % Volume LB | ratio mean % Volume ZnSO4 / mean % Volume LB | t-test mean % Volume ZnSO4 / mean % Volume LB |
| **Stringent response** | | | | | | | | | | | | |
| S1 | **YwaC** | P39583 | GTP pyrophosphokinase | 24600 | 6.2 | SigM, SigW | 2 | 14% | 7.2 | 96.5 | 5.8 | 99.7 |
| S2 | **SigH** | P17869 | RNA polymerase sigma-H factor | 25448 | 5.1 | AbrB | 7 | 47% | 0.6 | 98 | 0.6 | 99.8 |
| S3 | **YtxH** | P40780 | Uncharacterized protein | 16555 | 5.1 | SigB, SigH, stringent response | 4 | 32% | 1.7 | 98.1 | 1.3 | 51.6 |
| S4 | **YtzB** | O32065 | Uncharacterized protein | 11666 | 8.1 | stringent response | 3 | 52% | 0.6 | 97.4 | 0.7 | 90.5 |
| S5 | **TufA** | P33166 | Elongation factor Tu | 43593 | 4.7 | stringent response | 10 | 44% | 2.3 | 97.3 | 2.4 | 94.1 |
| S6 | **RplJ** | P42923 | 50S ribosomal protein L10 | 18078 | 5.5 | stringent response | 2 | 17% | 0.6 | 98.3 | 0.6 | 98.4 |
| S7 | **RpsB** | P21464 | 30S ribosomal protein S2 | 27968 | 6.2 | stringent response | 10 | 51% | 3.2 | 93.8 | 2.5 | 97.4 |
| S8 | **Mbl** | P39751 | MreB-like protein | 35861 | 5.7 | stringent response, sigE | 12 | 46% | 0.5 | 95.9 | 0.5 | 97.8 |
| **Metabolism** | | | | | | | | | | | | |
| **Cysteine metabolism** | | | | | | | | | | | | |
| C1 | **MetE** | P80877 | 5-methyltetrahydropteroyltriglutamate--homocysteine methyltransferase | 86808 | 4.8 | S-box | 17 | 31% | 0.4 | 99.8 | 0.6 | 87.6 |
| C2 | **MccB** | O05394 | Cystathionine gamma-lyase | 40886 | 5.2 | CymR , Spx | 10 | 44% | 1.5 | 97.6 | 2.1 | 98.0 |
| **central metabolism** | | | | | | | | | | | | |
| M1 | **OdhA** | P23129 | 2-oxoglutarate dehydrogenase E1 component | 106280 | 5.9 | CcpA | 31 | 43% | 0.6 | 92.9 | 0.5 | 98.0 |
| M2 | **AspB** | P53001 | Aspartate aminotransferase | 43089 | 5.3 | constitutive expression | 8 | 36% | 1.6 | 99.9 | 1.6 | 92.3 |
| M3 | **IolD** | P42415 | 3D-(3.5/4)-trihydroxycyclohexane-1.2-dione hydrolase | 70381 | 5.1 | CcpA, IolR | 17 | 42% | 0.6 | 99.4 | 0.6 | 97.3 |
| M4 | **AnsB** | P26899 | Aspartate ammonia-lyase | 52505 | 5.6 | AnsR | 7 | 29% | 2.0 | 97 | 1.3 | 79.2 |
| M5 | **PdhB** | P21882 | Pyruvate dehydrogenase E1 component subunit beta | 35474 | 4.5 | stringent response | 11 | 53% | 1.8 | 99.4 | 1.8 | 99.7 |
| M6 | **YtsJ** | O34962 | Probable NAD-dependent malic enzyme 4 | 43667 | 5.1 | unknown | 16 | 66% | 1.4 | 94.9 | 2.0 | 98.8 |
| **pentose pathway** | | | | | | | | | | | | |
| P1 | **XylA** | P0CI80 | Xylose isomerase | 50283 | 5.6 | CcpA, XylR | 11 | 36% | 0.5 | 97.6 | 0.4 | 98.7 |
| P2 | **YqjI** | P80859 | 6-phosphogluconate dehydrogenase. decarboxylating 2 | 51776 | 5.1 | Unknown | 20 | 67% | 2.5 | 99.8 | 0.9 | 60.2 |
| P3 | **Drm** | P46353 | Phosphopentomutase | 44020 | 4.9 | CcpA | 14 | 52% | 2.3 | 97.3 | 2.4 | 94.1 |
| P4 | **Tal** | P19669 | Transaldolase | 22972 | 5.9 | Unknown | 3 | 20% | 2.9 | 98.3 | 2.5 | 95.5 |
| **Other** | | | | | | | | | | | | |
| O1 | **FabL** | P71079 | Enoyl-[acyl-carrier-protein] reductase [NADPH] | 27178 | 5.9 | SigG, SpoVT, YfhP | 6 | 35% | 0.6 | 96.3 | 0.6 | 98.4 |
| O2 | **QcrA** | P46911 | Menaquinol-cytochrome c reductase iron-sulfur subunit | 18736 | 6.1 | AbrB, CcpA, ResD | 5 | 43% | 0.6 | 99 | 0.6 | 95.8 |
| O3 | **Nfo** | P54476 | Probable endonuclease 4 | 33069 | 5.4 | SigG | 6 | 34% | 1.6 | 99.8 | 1.2 | 80.3 |

**Table 1:** Differentially-expressed proteins identified in the proteomic screen. Comparison between 168 cytosolic crude extracts from 168 *Bacillus subtilis* strains grown in LB medium with or without stress (ZnO nanoparticles or $ZnSO_4$ salt stress) (Figure 2) (1) Spot number pointed out in Figure 2, (2) accession number from UniProKB, (3) protein function described by UniProtKB, (4) molecular weight, (5) pI, (6) number of peptides identified by mass spectrometry (7) percentage of coverage, (8) and (10) average quantification of the spot using Delta2D software from three independent 2-DE gels in the control, (9) and (11) standard deviation of the considering spot growth condition and (9) ratio of the average quantification determined by the Delta 2D analysis: stressed/not stressed.

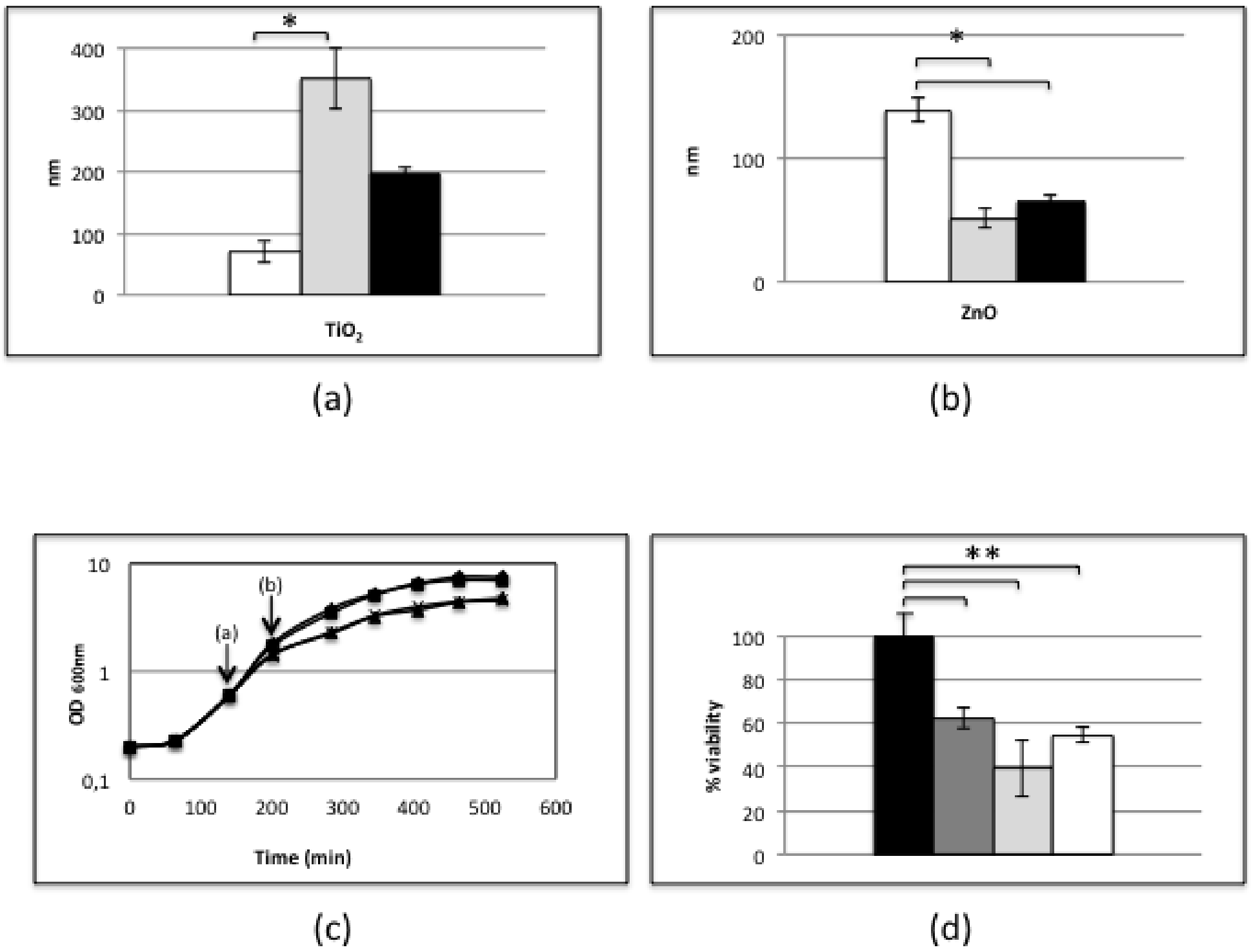

**Figure 1: DLS and viability measurements**

DLS measurements using a Wyatt Dynapro Nanostar instrument of (a) $TiO_2$ and (b) ZnO nanoparticles in different freshly prepared solutions: white for $H_2O$, grey for LB medium, black for conditioned LB medium (see Materials and Methods). (c) Viability measurements: all the assays were performed in triplicate with at least two technical replicates. The viability frequencies were normalized by the value obtained with the control. Normalized viability frequency of the control (LB) and the bacteria stressed with $TiO_2$, ZnO or $ZnSO_4$ are respectively represented in black, dark grey, soft grey and white histograms. Error bars ± 1SD ($n \geq 3$). Asterix * and ** indicate significant differences $p < 0.05$ and $p < 0.006$, respectively.

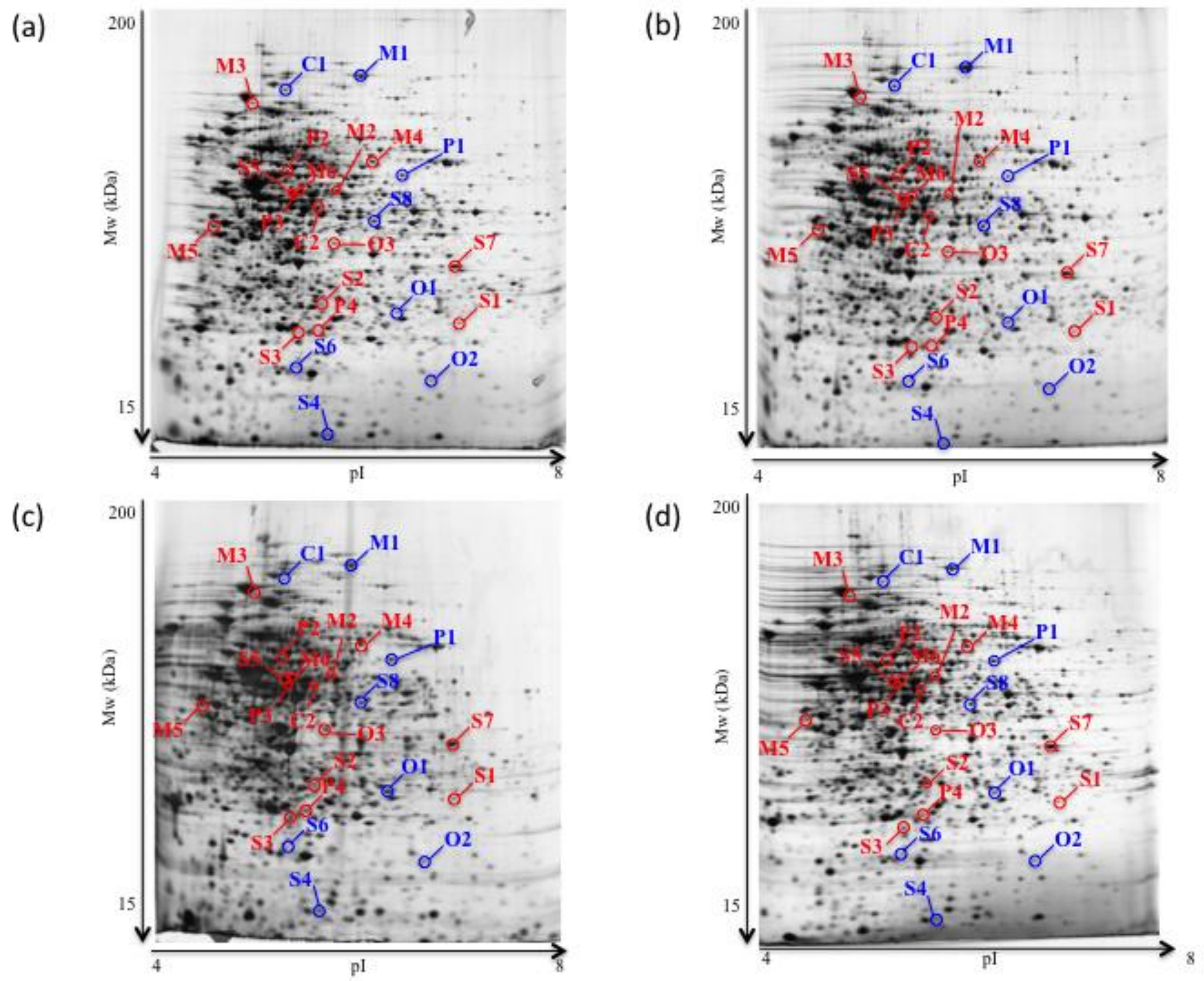

**Figure 2:** 2 DE gels of 168 *Bacillus subtilis* strain after ZnO or $ZnSO_4$ stresses

Total protein extracts of *Bacillus subtilis* 168 were separated by two-dimensional gel electrophoresis. The first dimensions covered a pH range of 4-8 and the second dimension was in the range of 15-200 kDa. Total cellular proteins (120 µg) were loaded on the first dimension gel, and the spots were detected by silver staining. (a) Gel obtained from the bacteria grown on LB medium, (b) gel obtained from bacteria stressed with 130 µg/ml $TiO_2$ for one hour, (c) gel obtained from bacteria stressed with 170 µg/ml ZnO or 0.3 mM $ZnSO_4$ for one hour and (d) gel obtained from bacteria stressed with 0.3 mM $ZnSO_4$ for one hour. The arrows point to spots that showed reproducible and statistically significant changes between the control conditions (LB) and the stress conditions (ZnO or $ZnSO_4$).

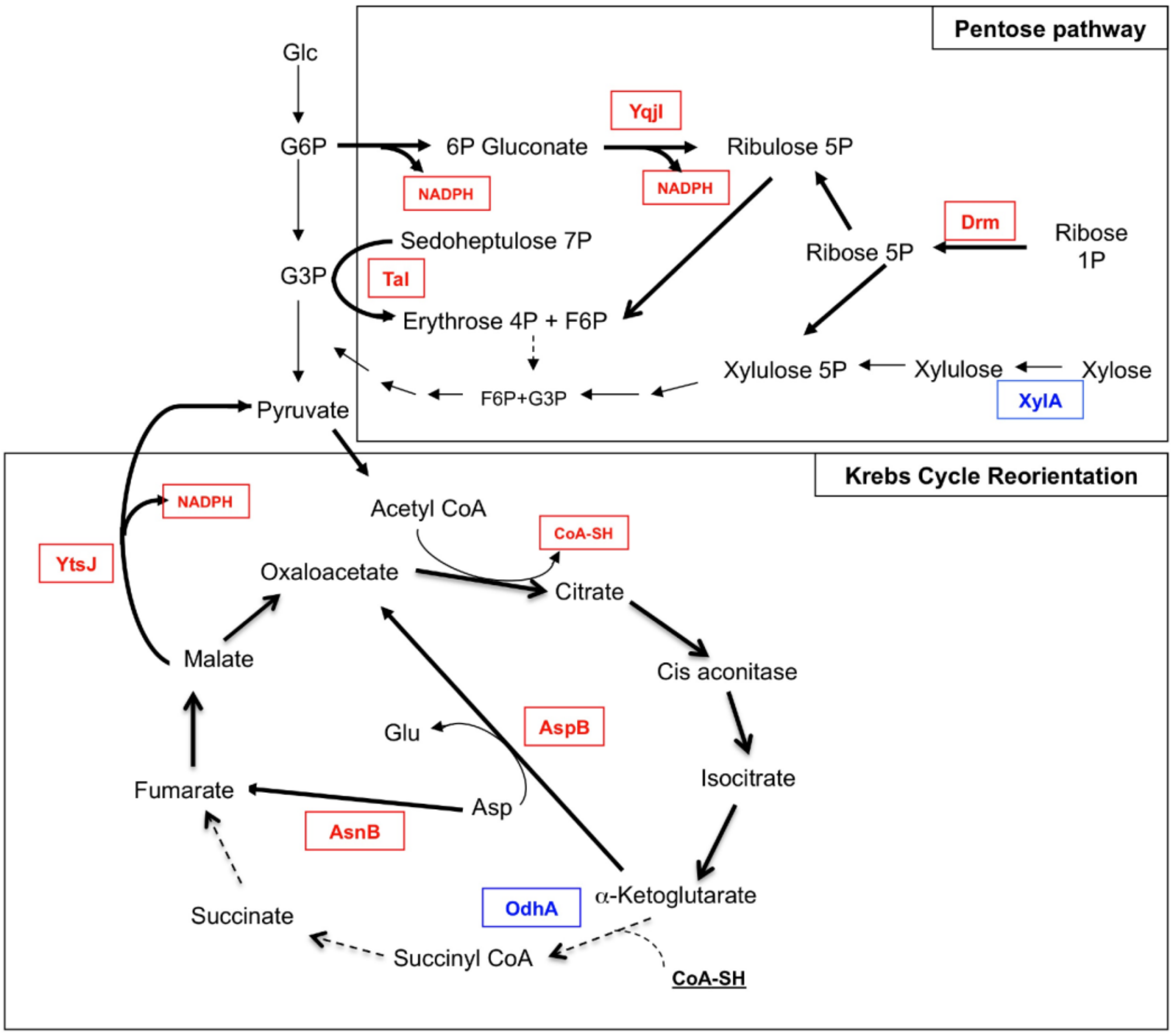

**Figure 3:** Metabolism reorientation

Enzymes highlighted in red and blue are proteins with an increase and decreased expression in presence of Zn stresses (see Table 1).

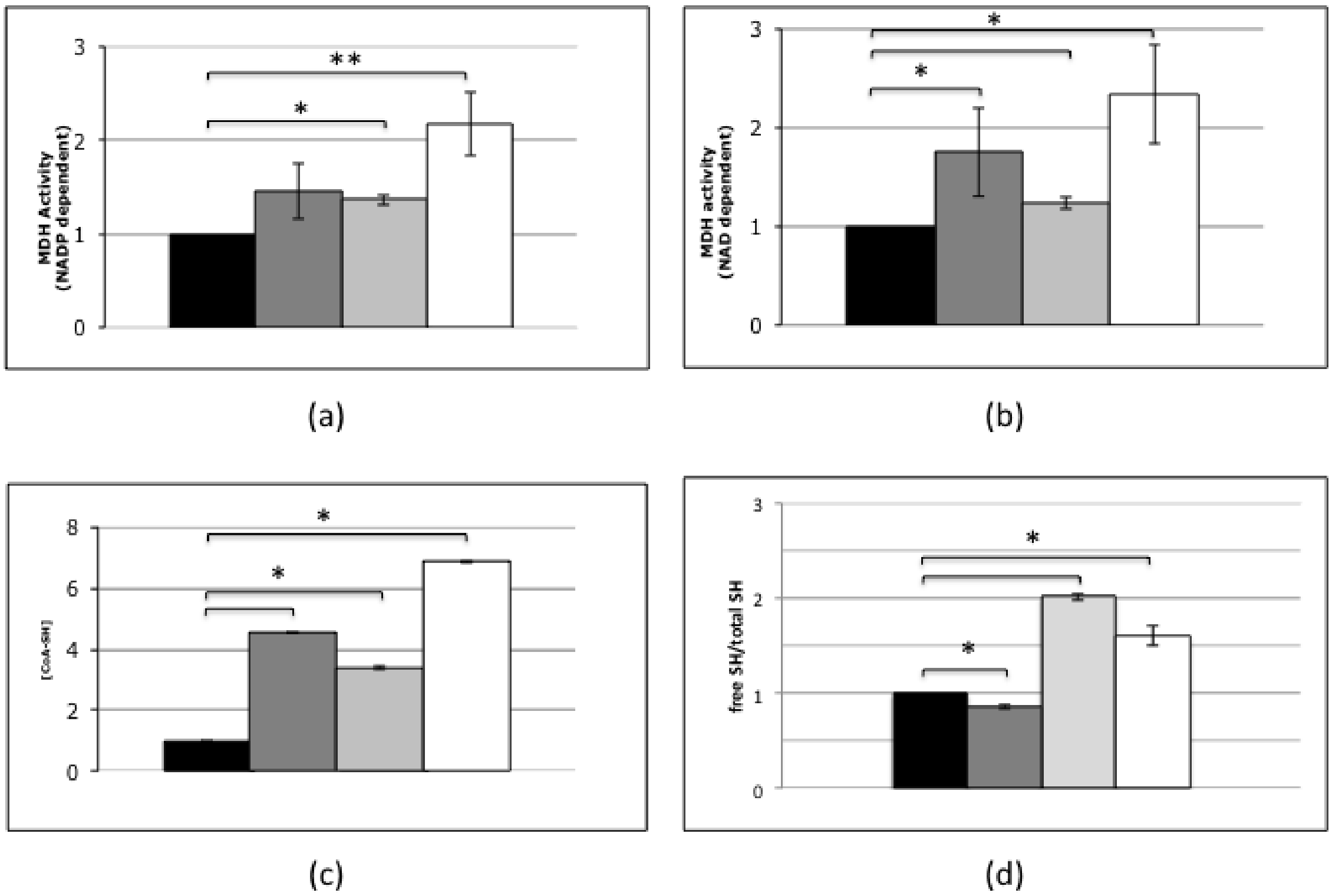

**Figure 4:** Malic enzyme activities (a) with NADP and (b) NAD as the cofactor. Histograms in (c) and (d) show the CoA assays and (free SH/(total SH-free SH)), respectively. All the assays were performed in triplicate with at least two technical replicates. The results were normalized by the value obtained with the control. Assays with the control bacteria (LB) and the bacteria stressed with $TiO_2$, ZnO or $ZnSO_4$ are respectively represented in black, dark grey, soft grey and white histograms. Error bars ± 1SD ($n \geq 3$). Asterix * and ** indicate significant differences $p < 0.05$ and $p < 0.006$, respectively.

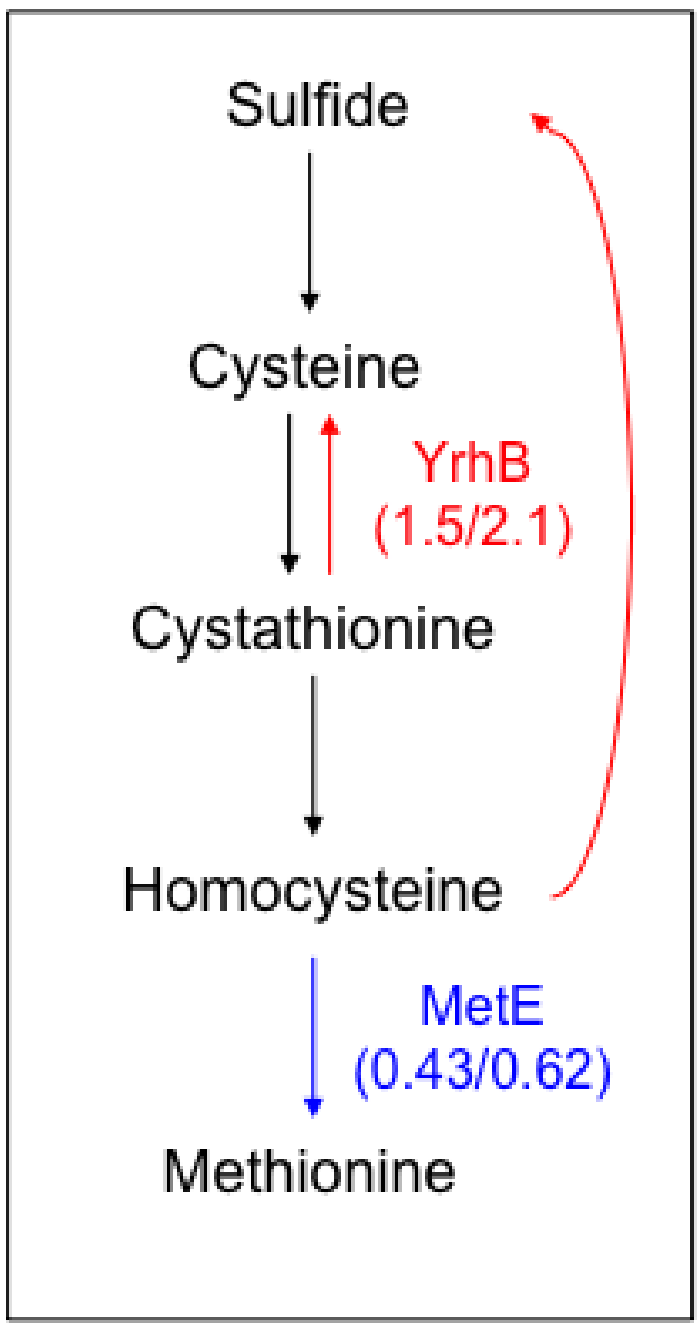

**Figure 5:** Cysteine-methionine metabolism

Values associated with YrhB and MetE refer to their expression modulations after a ZnO or $ZnSO_4$ stress compared to the (LB) control (Table 1).

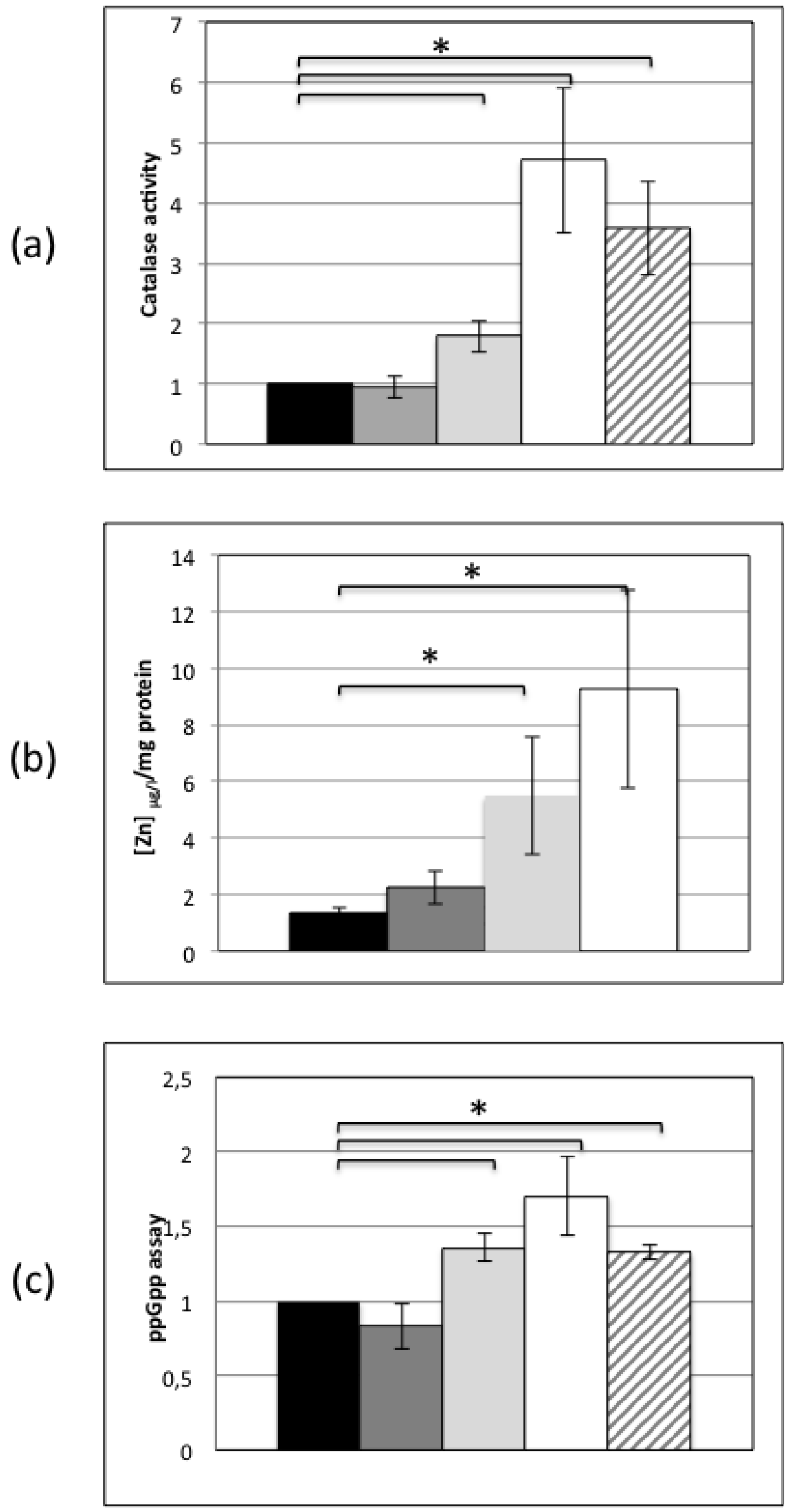

**Figure 6:** Histograms in (a), (b) and (c) show catalase activity, intracellular Zn concentration/mg protein and ppGpp assays, respectively. All the assays were performed in triplicate and at least two technical replicates. The results were normalized by the value obtained with the control (cells grown in LB medium at the same time as the stressed cells). Assays with the control bacteria (LB medium) and the bacteria stressed with $TiO_2$, ZnO, $ZnSO_4$ and (a) 5 mM $H_2O_2$ or (c) 1% norvaline are respectively represented in black, dark grey, light grey, white and dashed grey histograms. Error bars ± 1SD ($n \geq 3$). Asterix indicates significant differences $p < 0.05$.

**References:**

[1] Nel AE, Mädler L, Velegol D, Xia T, Hoek EMV, Somasundaran P, et al. Understanding biophysicochemical interactions at the nano-bio interface. Nat Mater 2009;8:543–57. doi:10.1038/nmat2442.

[2] Earl AM, Losick R, Kolter R. Ecology and genomics of Bacillus subtilis. Trends Microbiol 2008;16:269–75. doi:10.1016/j.tim.2008.03.004.

[3] Tareq FS, Kim JH, Lee MA, Lee H-S, Lee J-S, Lee Y-J, et al. Antimicrobial gageomacrolactins characterized from the fermentation of the marine-derived bacterium Bacillus subtilis under optimum growth conditions. J Agric Food Chem 2013;61:3428–34. doi:10.1021/jf4009229.

[4] Ran C, Carrias A, Williams MA, Capps N, Dan BCT, Newton JC, et al. Identification of Bacillus strains for biological control of catfish pathogens. PloS One 2012;7:e45793. doi:10.1371/journal.pone.0045793.

[5] Hecker M, Reder A, Fuchs S, Pagels M, Engelmann S. Physiological proteomics and stress/starvation responses in Bacillus subtilis and Staphylococcus aureus. Res Microbiol 2009;160:245–58. doi:10.1016/j.resmic.2009.03.008.

[6] Hahne H, Mäder U, Otto A, Bonn F, Steil L, Bremer E, et al. A comprehensive proteomics and transcriptomics analysis of Bacillus subtilis salt stress adaptation. J Bacteriol 2010;192:870–82. doi:10.1128/JB.01106-09.

[7] Zuber P. Management of oxidative stress in Bacillus. Annu Rev Microbiol 2009;63:575–97. doi:10.1146/annurev.micro.091208.073241.

[8] Hirooka K. Transcriptional response machineries of Bacillus subtilis conducive to plant growth promotion. Biosci Biotechnol Biochem 2014;78:1471–84. doi:10.1080/09168451.2014.943689.

[9] Vullo DL, Coto CE, Siñeriz F. Characteristics of an inulinase produced by Bacillus subtilis 430A, a strain isolated from the rhizosphere of Vernonia herbacea (Vell Rusby). Appl Environ Microbiol 1991;57:2392–4.

[10] Cazorla FM, Romero D, Pérez-García A, Lugtenberg BJJ, Vicente A de, Bloemberg G. Isolation and characterization of antagonistic Bacillus subtilis strains from the avocado rhizoplane displaying biocontrol activity. J Appl Microbiol 2007;103:1950–9. doi:10.1111/j.1365-2672.2007.03433.x.

[11] Bais HP, Fall R, Vivanco JM. Biocontrol of Bacillus subtilis against infection of Arabidopsis roots by Pseudomonas syringae is facilitated by biofilm formation and surfactin production. Plant Physiol 2004;134:307–19. doi:10.1104/pp.103.028712.

[12] Barbosa TM, Serra CR, La Ragione RM, Woodward MJ, Henriques AO. Screening for bacillus isolates in the broiler gastrointestinal tract. Appl Environ Microbiol 2005;71:968–78. doi:10.1128/AEM.71.2.968-978.2005.

[13] Tam NKM, Uyen NQ, Hong HA, Duc LH, Hoa TT, Serra CR, et al. The intestinal life cycle of Bacillus subtilis and close relatives. J Bacteriol 2006;188:2692–700. doi:10.1128/JB.188.7.2692-2700.2006.

[14] Serra CR, Earl AM, Barbosa TM, Kolter R, Henriques AO. Sporulation during growth in a gut isolate of Bacillus subtilis. J Bacteriol 2014;196:4184–96. doi:10.1128/JB.01993-14.

[15] Sirec T, Cangiano G, Baccigalupi L, Ricca E, Isticato R. The spore surface of intestinal isolates of Bacillus subtilis. FEMS Microbiol Lett 2014;358:194–201. doi:10.1111/1574-6968.12538.

[16] Adams LK, Lyon DY, McIntosh A, Alvarez PJJ. Comparative toxicity of nano-scale TiO2, SiO2 and ZnO water suspensions. Water Sci Technol J Int Assoc Water Pollut Res 2006;54:327–34.

[17] Adams LK, Lyon DY, Alvarez PJJ. Comparative eco-toxicity of nanoscale TiO2,

SiO2, and ZnO water suspensions. Water Res 2006;40:3527–32. doi:10.1016/j.watres.2006.08.004.
[18] Azam A, Ahmed AS, Oves M, Khan MS, Habib SS, Memic A. Antimicrobial activity of metal oxide nanoparticles against Gram-positive and Gram-negative bacteria: a comparative study. Int J Nanomedicine 2012;7:6003–9. doi:10.2147/IJN.S35347.
[19] Baek Y-W, An Y-J. Microbial toxicity of metal oxide nanoparticles (CuO, NiO, ZnO, and Sb2O3) to Escherichia coli, Bacillus subtilis, and Streptococcus aureus. Sci Total Environ 2011;409:1603–8. doi:10.1016/j.scitotenv.2011.01.014.
[20] Fang J, Lyon DY, Wiesner MR, Dong J, Alvarez PJJ. Effect of a fullerene water suspension on bacterial phospholipids and membrane phase behavior. Environ Sci Technol 2007;41:2636–42.
[21] Kim SW, An Y-J. Effect of ZnO and $TiO_2$ nanoparticles preilluminated with UVA and UVB light on Escherichia coli and Bacillus subtilis. Appl Microbiol Biotechnol 2012;95:243–53. doi:10.1007/s00253-012-4153-6.
[22] Lyon DY, Brunet L, Hinkal GW, Wiesner MR, Alvarez PJJ. Antibacterial activity of fullerene water suspensions (nC60) is not due to ROS-mediated damage. Nano Lett 2008;8:1539–43. doi:10.1021/nl0726398.
[23] Lyon DY, Alvarez PJJ. Fullerene water suspension (nC60) exerts antibacterial effects via ROS-independent protein oxidation. Environ Sci Technol 2008;42:8127–32.
[24] Raliya R, Tarafdar JC, Mahawar H, Kumar R, Gupta P, Mathur T, et al. ZnO nanoparticles induced exopolysaccharide production by B. subtilis strain JCT1 for arid soil applications. Int J Biol Macromol 2014;65:362–8. doi:10.1016/j.ijbiomac.2014.01.060.
[25] Santimano MC, Kowshik M. Altered growth and enzyme expression profile of ZnO nanoparticles exposed non-target environmentally beneficial bacteria. Environ Monit Assess 2013;185:7205–14. doi:10.1007/s10661-013-3094-6.
[26] Triboulet S, Aude-Garcia C, Armand L, Gerdil A, Diemer H, Proamer F, et al. Analysis of cellular responses of macrophages to zinc ions and zinc oxide nanoparticles: a combined targeted and proteomic approach. Nanoscale 2014;6:6102–14. doi:10.1039/c4nr00319e.
[27] Triboulet S, Aude-Garcia C, Armand L, Collin-Faure V, Chevallet M, Diemer H, et al. Comparative proteomic analysis of the molecular responses of mouse macrophages to titanium dioxide and copper oxide nanoparticles unravels some toxic mechanisms for copper oxide nanoparticles in macrophages. PloS One 2015.
[28] Luche S, Diemer H, Tastet C, Chevallet M, Van Dorsselaer A, Leize-Wagner E, et al. About thiol derivatization and resolution of basic proteins in two-dimensional electrophoresis. Proteomics 2004;4:551–61. doi:10.1002/pmic.200300589.
[29] Rabilloud T, Adessi C, Giraudel A, Lunardi J. Improvement of the solubilization of proteins in two-dimensional electrophoresis with immobilized pH gradients. Electrophoresis 1997;18:307–16. doi:10.1002/elps.1150180303.
[30] Neuhoff V, Arold N, Taube D, Ehrhardt W. Improved staining of proteins in polyacrylamide gels including isoelectric focusing gels with clear background at nanogram sensitivity using Coomassie Brilliant Blue G-250 and R-250. Electrophoresis 1988;9:255–62. doi:10.1002/elps.1150090603.
[31] Gianazza E, Celentano F, Magenes S, Ettori C, Righetti PG. Formulations for immobilized pH gradients including pH extremes. Electrophoresis 1989;10:806–8. doi:10.1002/elps.1150101115.
[32] Rabilloud T, Valette C, Lawrence JJ. Sample application by in-gel rehydration improves the resolution of two-dimensional electrophoresis with immobilized pH gradients in the first dimension. Electrophoresis 1994;15:1552–8.
[33] Tastet C, Lescuyer P, Diemer H, Luche S, van Dorsselaer A, Rabilloud T. A versatile

electrophoresis system for the analysis of high- and low-molecular-weight proteins. Electrophoresis 2003;24:1787–94. doi:10.1002/elps.200305400.
[34] Chevallet M, Luche S, Rabilloud T. Silver staining of proteins in polyacrylamide gels. Nat Protoc 2006;1:1852–8. doi:10.1038/nprot.2006.288.
[35] Storey JD, Tibshirani R. Statistical significance for genomewide studies. Proc Natl Acad Sci U S A 2003;100:9440–5. doi:10.1073/pnas.1530509100.
[36] Karp NA, McCormick PS, Russell MR, Lilley KS. Experimental and statistical considerations to avoid false conclusions in proteomics studies using differential in-gel electrophoresis. Mol Cell Proteomics MCP 2007;6:1354–64. doi:10.1074/mcp.M600274-MCP200.
[37] Gharahdaghi F, Weinberg CR, Meagher DA, Imai BS, Mische SM. Mass spectrometric identification of proteins from silver-stained polyacrylamide gel: a method for the removal of silver ions to enhance sensitivity. Electrophoresis 1999;20:601–5. doi:10.1002/(SICI)1522-2683(19990301)20:3<601::AID-ELPS601>3.0.CO;2-6.
[38] Richert S, Luche S, Chevallet M, Van Dorsselaer A, Leize-Wagner E, Rabilloud T. About the mechanism of interference of silver staining with peptide mass spectrometry. Proteomics 2004;4:909–16. doi:10.1002/pmic.200300642.
[39] Triboulet S, Aude-Garcia C, Carrière M, Diemer H, Proamer F, Habert A, et al. Molecular responses of mouse macrophages to copper and copper oxide nanoparticles inferred from proteomic analyses. Mol Cell Proteomics MCP 2013;12:3108–22. doi:10.1074/mcp.M113.030742.
[40] Lerondel G, Doan T, Zamboni N, Sauer U, Aymerich S. YtsJ has the major physiological role of the four paralogous malic enzyme isoforms in Bacillus subtilis. J Bacteriol 2006;188:4727–36. doi:10.1128/JB.00167-06.
[41] Rorth M, Jensen PK. Determination of catalase activity by means of the Clark oxygen electrode. Biochim Biophys Acta 1967;139:171–3.
[42] Zhang P, Wang Y, Chang Y, Xiong ZH, Huang CZ. Highly selective detection of bacterial alarmone ppGpp with an off-on fluorescent probe of copper-mediated silver nanoclusters. Biosens Bioelectron 2013;49:433–7. doi:10.1016/j.bios.2013.05.056.
[43] Li M, Zhu L, Lin D. Toxicity of ZnO nanoparticles to Escherichia coli: mechanism and the influence of medium components. Environ Sci Technol 2011;45:1977–83. doi:10.1021/es102624t.
[44] Reitzer LJ, Wice BM, Kennell D. The pentose cycle. Control and essential function in HeLa cell nucleic acid synthesis. J Biol Chem 1980;255:5616–26.
[45] Zamboni N, Fischer E, Laudert D, Aymerich S, Hohmann H-P, Sauer U. The Bacillus subtilis yqjI gene encodes the NADP+-dependent 6-P-gluconate dehydrogenase in the pentose phosphate pathway. J Bacteriol 2004;186:4528–34. doi:10.1128/JB.186.14.4528-4534.2004.
[46] Meyer FM, Stülke J. Malate metabolism in Bacillus subtilis: distinct roles for three classes of malate-oxidizing enzymes. FEMS Microbiol Lett 2013;339:17–22. doi:10.1111/1574-6968.12041.
[47] Ma Z, Chandrangsu P, Helmann TC, Romsang A, Gaballa A, Helmann JD. Bacillithiol is a major buffer of the labile zinc pool in Bacillus subtilis. Mol Microbiol 2014;94:756–70. doi:10.1111/mmi.12794.
[48] Posada AC, Kolar SL, Dusi RG, Francois P, Roberts AA, Hamilton CJ, et al. Importance of Bacillithiol in the Oxidative Stress Response of Staphylococcus aureus. Infect Immun 2014;82:316–32. doi:10.1128/IAI.01074-13.
[49] Chi BK, Gronau K, Mäder U, Hessling B, Becher D, Antelmann H. S-bacillithiolation protects against hypochlorite stress in Bacillus subtilis as revealed by transcriptomics and redox proteomics. Mol Cell Proteomics MCP 2011;10:M111.009506. doi:10.1074/mcp.M111.009506.

[50] Gaballa A, Newton GL, Antelmann H, Parsonage D, Upton H, Rawat M, et al. Biosynthesis and functions of bacillithiol, a major low-molecular-weight thiol in Bacilli. Proc Natl Acad Sci U S A 2010;107:6482–6. doi:10.1073/pnas.1000928107.
[51] Helmann JD. Bacillithiol, a new player in bacterial redox homeostasis. Antioxid Redox Signal 2011;15:123–33. doi:10.1089/ars.2010.3562.
[52] Lee J-W, Soonsanga S, Helmann JD. A complex thiolate switch regulates the Bacillus subtilis organic peroxide sensor OhrR. Proc Natl Acad Sci U S A 2007;104:8743–8. doi:10.1073/pnas.0702081104.
[53] Tam LT, Antelmann H, Eymann C, Albrecht D, Bernhardt J, Hecker M. Proteome signatures for stress and starvation in Bacillus subtilis as revealed by a 2-D gel image color coding approach. Proteomics 2006;6:4565–85. doi:10.1002/pmic.200600100.
[54] Eymann C, Mittenhuber G, Hecker M. The stringent response, sigmaH-dependent gene expression and sporulation in Bacillus subtilis. Mol Gen Genet MGG 2001;264:913–23.
[55] Nanamiya H, Kasai K, Nozawa A, Yun C-S, Narisawa T, Murakami K, et al. Identification and functional analysis of novel (p)ppGpp synthetase genes in Bacillus subtilis. Mol Microbiol 2008;67:291–304. doi:10.1111/j.1365-2958.2007.06018.x.
[56] Natori Y, Tagami K, Murakami K, Yoshida S, Tanigawa O, Moh Y, et al. Transcription activity of individual rrn operons in Bacillus subtilis mutants deficient in (p)ppGpp synthetase genes, relA, yjbM, and ywaC. J Bacteriol 2009;191:4555–61. doi:10.1128/JB.00263-09.
[57] Tagami K, Nanamiya H, Kazo Y, Maehashi M, Suzuki S, Namba E, et al. Expression of a small (p)ppGpp synthetase, YwaC, in the (p)ppGpp(0) mutant of Bacillus subtilis triggers YvyD-dependent dimerization of ribosome. MicrobiologyOpen 2012;1:115–34. doi:10.1002/mbo3.16.
[58] Kriel A, Brinsmade SR, Tse JL, Tehranchi AK, Bittner AN, Sonenshein AL, et al. GTP dysregulation in Bacillus subtilis cells lacking (p)ppGpp results in phenotypic amino acid auxotrophy and failure to adapt to nutrient downshift and regulate biosynthesis genes. J Bacteriol 2014;196:189–201. doi:10.1128/JB.00918-13.
[59] Hochgräfe F, Mostertz J, Pöther D-C, Becher D, Helmann JD, Hecker M. S-cysteinylation is a general mechanism for thiol protection of Bacillus subtilis proteins after oxidative stress. J Biol Chem 2007;282:25981–5. doi:10.1074/jbc.C700105200.
[60] Maret W, Li Y. Coordination dynamics of zinc in proteins. Chem Rev 2009;109:4682–707. doi:10.1021/cr800556u.
[61] Maret W. Inhibitory zinc sites in enzymes. Biometals Int J Role Met Ions Biol Biochem Med 2013;26:197–204. doi:10.1007/s10534-013-9613-7.
[62] Wouters MA, Iismaa S, Fan SW, Haworth NL. Thiol-based redox signalling: rust never sleeps. Int J Biochem Cell Biol 2011;43:1079–85. doi:10.1016/j.biocel.2011.04.002.
[63] Lemire J, Mailloux R, Appanna VD. Zinc toxicity alters mitochondrial metabolism and leads to decreased ATP production in hepatocytes. J Appl Toxicol JAT 2008;28:175–82. doi:10.1002/jat.1263.
[64] Guirola M, Jiménez-Martí E, Atrian S. On the molecular relationships between high-zinc tolerance and aconitase (Aco1) in Saccharomyces cerevisiae. Met Integr Biometal Sci 2014;6:634–45. doi:10.1039/c3mt00360d.
[65] Kasahara M, Anraku Y. Succinate- and NADH oxidase systems of Escherichia coli membrane vesicles. Mechanism of selective inhibition of the systems by zinc ions. J Biochem (Tokyo) 1974;76:967–76.
[66] Sigdel TK, Cilliers R, Gursahaney PR, Thompson P, Easton JA, Crowder MW. Probing the adaptive response of Escherichia coli to extracellular Zn(II). Biometals Int J Role Met Ions Biol Biochem Med 2006;19:461–71. doi:10.1007/s10534-005-4962-5.
[67] Eide DJ. Bacillithiol, a new role in buffering intracellular zinc. Mol Microbiol

2014;94:743–6. doi:10.1111/mmi.12793.
[68] Spira B, Hu X, Ferenci T. Strain variation in ppGpp concentration and RpoS levels in laboratory strains of Escherichia coli K-12. Microbiol Read Engl 2008;154:2887–95. doi:10.1099/mic.0.2008/018457-0.
[69] Eymann C, Homuth G, Scharf C, Hecker M. Bacillus subtilis functional genomics: global characterization of the stringent response by proteome and transcriptome analysis. J Bacteriol 2002;184:2500–20.
[70] Petersohn A, Brigulla M, Haas S, Hoheisel JD, Völker U, Hecker M. Global analysis of the general stress response of Bacillus subtilis. J Bacteriol 2001;183:5617–31. doi:10.1128/JB.183.19.5617-5631.2001.
[71] Bowden SD, Eyres A, Chung JCS, Monson RE, Thompson A, Salmond GPC, et al. Virulence in Pectobacterium atrosepticum is regulated by a coincidence circuit involving quorum sensing and the stress alarmone, (p)ppGpp. Mol Microbiol 2013;90:457–71. doi:10.1111/mmi.12369.
[72] Nowicki D, Maciąg-Dorszyńska M, Kobiela W, Herman-Antosiewicz A, Węgrzyn A, Szalewska-Pałasz A, et al. Phenethyl isothiocyanate inhibits shiga toxin production in enterohemorrhagic Escherichia coli by stringent response induction. Antimicrob Agents Chemother 2014;58:2304–15. doi:10.1128/AAC.02515-13.
[73] Oh YT, Park Y, Yoon MY, Bari W, Go J, Min KB, et al. Cholera toxin production during anaerobic trimethylamine N-oxide respiration is mediated by stringent response in Vibrio cholerae. J Biol Chem 2014;289:13232–42. doi:10.1074/jbc.M113.540088.
[74] Tabone M, Lioy VS, Ayora S, Machón C, Alonso JC. Role of toxin ζ and starvation responses in the sensitivity to antimicrobials. PloS One 2014;9:e86615. doi:10.1371/journal.pone.0086615.
[75] Ochi K, Kandala J, Freese E. Evidence that Bacillus subtilis sporulation induced by the stringent response is caused by the decrease in GTP or GDP. J Bacteriol 1982;151:1062–5.
[76] Mirouze N, Prepiak P, Dubnau D. Fluctuations in spo0A transcription control rare developmental transitions in Bacillus subtilis. PLoS Genet 2011;7:e1002048. doi:10.1371/journal.pgen.1002048.
[77] Geiger T, Wolz C. Intersection of the stringent response and the CodY regulon in low GC Gram-positive bacteria. Int J Med Microbiol IJMM 2014;304:150–5. doi:10.1016/j.ijmm.2013.11.013.
[78] Geiger T, Francois P, Liebeke M, Fraunholz M, Goerke C, Krismer B, et al. The stringent response of Staphylococcus aureus and its impact on survival after phagocytosis through the induction of intracellular PSMs expression. PLoS Pathog 2012;8:e1003016. doi:10.1371/journal.ppat.1003016.
[79] Lemos JA, Nascimento MM, Lin VK, Abranches J, Burne RA. Global regulation by (p)ppGpp and CodY in Streptococcus mutans. J Bacteriol 2008;190:5291–9. doi:10.1128/JB.00288-08.
[80] Schoenfelder SMK, Marincola G, Geiger T, Goerke C, Wolz C, Ziebuhr W. Methionine biosynthesis in Staphylococcus aureus is tightly controlled by a hierarchical network involving an initiator tRNA-specific T-box riboswitch. PLoS Pathog 2013;9:e1003606. doi:10.1371/journal.ppat.1003606.
[81] Hsueh Y-H, Ke W-J, Hsieh C-T, Lin K-S, Tzou D-Y, Chiang C-L. ZnO Nanoparticles Affect Bacillus subtilis Cell Growth and Biofilm Formation. PloS One 2015;10:e0128457. doi:10.1371/journal.pone.0128457.
[82] Marvasi M, Visscher PT, Casillas Martinez L. Exopolymeric substances (EPS) from Bacillus subtilis: polymers and genes encoding their synthesis. FEMS Microbiol Lett 2010;313:1–9. doi:10.1111/j.1574-6968.2010.02085.x.

[83] Notter DA, Mitrano DM, Nowack B. Are nanosized or dissolved metals more toxic in the environment? A meta-analysis. Environ Toxicol Chem SETAC 2014;33:2733–9. doi:10.1002/etc.2732.
[84] Gottschalk F, Sun T, Nowack B. Environmental concentrations of engineered nanomaterials: review of modeling and analytical studies. Environ Pollut Barking Essex 1987 2013;181:287–300. doi:10.1016/j.envpol.2013.06.003.
[85] Vittori Antisari L, Carbone S, Gatti A, Vianello G, Nannipieri P. Uptake and translocation of metals and nutrients in tomato grown in soil polluted with metal oxide (CeO2, Fe 3O 4, SnO 2, TiO 2) or metallic (Ag, Co, Ni) engineered nanoparticles. Environ Sci Pollut Res Int 2015;22:1841–53. doi:10.1007/s11356-014-3509-0.
[86] Thiruvengadam M, Gurunathan S, Chung I-M. Physiological, metabolic, and transcriptional effects of biologically-synthesized silver nanoparticles in turnip (Brassica rapa ssp. rapa L.). Protoplasma 2014. doi:10.1007/s00709-014-0738-5.
[87] Larue C, Laurette J, Herlin-Boime N, Khodja H, Fayard B, Flank A-M, et al. Accumulation, translocation and impact of TiO2 nanoparticles in wheat (Triticum aestivum spp.): influence of diameter and crystal phase. Sci Total Environ 2012;431:197–208. doi:10.1016/j.scitotenv.2012.04.073.
[88] Larue C, Castillo-Michel H, Sobanska S, Trcera N, Sorieul S, Cécillon L, et al. Fate of pristine TiO2 nanoparticles and aged paint-containing TiO2 nanoparticles in lettuce crop after foliar exposure. J Hazard Mater 2014;273:17–26. doi:10.1016/j.jhazmat.2014.03.014.
[89] Bandyopadhyay S, Plascencia-Villa G, Mukherjee A, Rico CM, José-Yacamán M, Peralta-Videa JR, et al. Comparative phytotoxicity of ZnO NPs, bulk ZnO, and ionic zinc onto the alfalfa plants symbiotically associated with Sinorhizobium meliloti in soil. Sci Total Environ 2015;515-516C:60–9. doi:10.1016/j.scitotenv.2015.02.014.
[90] Dimkpa CO. Can nanotechnology deliver the promised benefits without negatively impacting soil microbial life? J Basic Microbiol 2014;54:889–904. doi:10.1002/jobm.201400298.
[91] Dimkpa CO, McLean JE, Britt DW, Anderson AJ. Nano-CuO and interaction with nano-ZnO or soil bacterium provide evidence for the interference of nanoparticles in metal nutrition of plants. Ecotoxicol Lond Engl 2015;24:119–29. doi:10.1007/s10646-014-1364-x.
[92] Fan R, Huang YC, Grusak MA, Huang CP, Sherrier DJ. Effects of nano-$TiO_2$ on the agronomically-relevant Rhizobium-legume symbiosis. Sci Total Environ 2014;466-467:503–12. doi:10.1016/j.scitotenv.2013.07.032.
[93] Maurer-Jones MA, Gunsolus IL, Murphy CJ, Haynes CL. Toxicity of engineered nanoparticles in the environment. Anal Chem 2013;85:3036–49. doi:10.1021/ac303636s.